\documentclass[aps, twocolumn, prb, floatfix, superscriptaddress,longbibliography]{revtex4-2}
\usepackage{amsmath,amssymb,amsthm}
\usepackage{graphicx}
\usepackage{color}
\definecolor{darkblue}{rgb}{0,0,.65}
\definecolor{darkgreen}{rgb}{0.3,0.6,0.3}
\definecolor{cyan1}{rgb}{0.0, 0.6, 0.6}
\usepackage[%
pdfstartview=FitH,%
breaklinks=true,%
bookmarks=true,%
colorlinks=true,%
anchorcolor=black,%
citecolor=red,
filecolor=black,%
menucolor=black,%
urlcolor=darkblue,%
linkcolor=blue,%
]{hyperref}
\usepackage{cleveref}

\definecolor{myblue}{rgb}{0,0,0.75}

\newcommand{\be}{\begin{equation}}
\newcommand{\ee}{\end{equation}}
\def\ba{\begin{aligned}}
\def\ea{\end{aligned}}
\newcommand{\bea}{\begin{eqnarray}}
\newcommand{\eea}{\end{eqnarray}}
\def\bes{\begin{subequations}}
\def\ees{\end{subequations}}
\def\bal{\begin{align}}
\def\eal{\end{align}}

\usepackage{soul}
\usepackage{ulem} % needed for \sout-command

\usepackage{graphicx}
\usepackage{amsmath,amssymb,bm}
\usepackage{enumerate}
\usepackage{braket}        % Dirac notation

\begin{document}

\title{Aperiodic Dissipation as a Mechanism for Steady-State Localization}

\author{Shilpi Roy}
\email{roy_s@nus.edu.sg}
\affiliation{Department of Physics, National University of Singapore, Singapore 117542, Singapore}

\author{Jiangbin Gong}
\email{phygj@nus.edu.sg}

\affiliation{Department of Physics, National University of Singapore, Singapore 117542, Singapore}
\affiliation{Centre for Quantum Technologies, National University of Singapore, Singapore 117543, Singapore}
\affiliation{MajuLab, CNRS-UCA-SU-NUS-NTU International Joint Research Unit, Singapore}

%\author{}
%\affiliation{}

% %\date{\today}%
\begin{abstract}
Dissipation is traditionally regarded as a disruptive factor in quantum systems because it often leads to decoherence and delocalization. However, recent insights into engineered dissipation reveal that it can be tuned to facilitate various quantum effects, from state stabilization to phase transitions. In this work, we identify aperiodic dissipation as a mechanism for inducing steady-state localization, independent of disorder or a quasiperiodic potential in the Hamiltonian. This localization arises from long-range phase correlations introduced by a spatially varying dissipation phase parameter, which enables nontrivial interference in the steady-state.
By systematically comparing two classes of aperiodic dissipation (defined as commensurate and incommensurate cases), we find that incommensurate modulation plays the most efficient role in stabilizing a localized steady-state. Our analysis, based on coherence measures, purity, and participation ratio, reveals a direct link between eigenstate coherence and real-space localization, showing that dissipation can actively shape localization rather than simply causing decoherence. These findings highlight aperiodic dissipation as a viable approach to controlling localization in open quantum systems, potentially enabling new ways to manipulate quantum states and design dissipation-driven phases.
\end{abstract}

\maketitle

\section{Introduction}
Dissipation is an unavoidable feature of any quantum system since complete isolation from the environment is fundamentally unattainable. When a system interacts with its surroundings, it undergoes decoherence~\cite{baumgratz2014quantifying, nielsen2010quantum}. As a result, important quantum phenomenon such as Anderson localization~\cite{anderson1958absence, lagendijk2009fifty} and quantum entanglement~\cite{plenio2014introduction, vedral1998entanglement}, which rely on interference effects, are typically disrupted by dissipation. Nevertheless, dissipation is not always detrimental. When carefully engineered, it can be harnessed as a powerful tool for controlling quantum states rather than merely destroying them.  

The field of engineered dissipation has fundamentally reshaped our understanding of open quantum systems, revealing that dissipation can be actively tailored to induce specific quantum effects~\cite{harrington2022engineered, damanet2019controlling, shah2024stabilizing}. This has led to remarkable possibilities, including the preparation of desirable pure~\cite{diehl2008quantum} and entangled states~\cite{kraus2008preparation, verstraete2009quantum, irfan2024loss}, the stabilization of metastable states~\cite{valenti2015stabilization, magazzu2015dissipative}, and even the emergence of coherence~\cite{syassen2008strong, witthaut2008dissipation}. Dissipation has also been recognized as a driving force behind phase transitions~\cite{prosen2008quantum, mebrahtu2012quantum,medvedyeva2015dissipation, shastri2020dissipation, yamamoto2021collective, soriente2021distinctive, zhou2024nonreciprocal}.  This growing understanding has fueled significant interest in dissipation-induced localization and transport phenomena, particularly in disordered~\cite{gurvitz2000delocalization, huse2015localized, yusipov2017localization, vershinina2017control, yusipov2018quantum, balasubrahmaniyam2020necklace, weidemann2021coexistence, rath2020prominent, hunter2020quantum,longhi2023anderson}, quasiperiodic quantum systems~\cite{purkayastha2017nonequilibrium, balachandran2019energy, chiaracane2020quasiperiodic, lacerda2021dephasing, dwiputra2021environment, chiaracane2022dephasing, saha2022quantum, liu2024dissipation, longhi2024dephasing}, as well as their many-body counterparts~\cite{fischer2016dynamics, levi2016robustness, everest2017role, luschen2017signatures, vakulchyk2018signatures, xu2018interplay, wu2019bath, wybo2020entanglement, lenarvcivc2020critical, gupta2024quantum}. These advances establish dissipation not just as an obstacle, but as a fundamental element in quantum control, opening exciting new directions in both theoretical and experimental research.

A key advancement in this field has been the realization that phase-modulated dissipation can serve as a control knob, selectively stabilizing certain eigenstates over others. This concept was demonstrated in Ref.~\cite{yusipov2017localization}, where a phase-modulated dissipation operator was shown to stabilize Anderson-localized states in the steady-state. By shielding specific eigenstates from dissipation, the system preserves localization even in the presence of an environment.
More recently, Ref.~\cite{liu2024dissipation} investigated the same dissipation operator in a system with mobility edges, where both extended and localized states coexist within the Hamiltonian. It was found that dissipation could selectively stabilize either extended or localized states depending on the phase parameter, thereby inducing a dissipation-driven transition between extended and localized steady-states.

These studies establish a fundamental connection between the localization properties of Hamiltonian eigenstates and the emergence of steady-state localization under engineered dissipation~\cite{yusipov2017localization, liu2024dissipation}. They suggest the following scenario: steady-state localization can occur when the eigenstates of the Hamiltonian, which are favored by dissipation, are already localized. This link or understanding motivates a fundamental and curious question: Can dissipation alone induce localization in an otherwise clean, extended system?
Building upon the above-mentioned early studies, we give an affirmative answer below and also provide physics perspectives beyond what is known so far.

In this work, we explore whether site-dependent, phase-modulated dissipation can independently induce localization in a clean, extended quantum system. Specifically, we investigate whether dissipation alone can induce steady-state localization, without requiring disorder or a quasiperiodic potential in the Hamiltonian. To address this, we consider aperiodic dissipation, which inherently induces long-range phase correlations and plays a crucial role in shaping steady-state localization~\cite{sarma1988mobility, sarma1990localization}. We systematically compare two classes of this aperiodic dissipation (defined to be commensurate and incommensurate cases), by examining how phase modulation influences steady-state localization.
Our extensive numerical simulations enable us to investigate key observables, such as the relative entropy of coherence, purity, and participation ratio, to characterize the interplay between dissipation, coherence retention, and spatial localization. Through this analysis, we establish aperiodic dissipation as an effective mechanism for engineered localization, a piece of physics distinct from all previously known possibilities of disorder-driven localization.

The paper is organized as follows: Section~\ref{Model} introduces the model Hamiltonian of our  quantum lattice system under investigation, the dissipation operator, and the implementation of phase modulation in our model. Section~\ref{Results} presents our results, where we systematically analyze the effects of \ref{Incommensurate_case}~aperiodic incommensurate and \ref{Commensurate_case}~aperiodic commensurate dissipation, highlighting their differences in steady-state localization. In Section~\ref{Experimental_Proposal}, we outline a possible experimental implementation of our dissipation mechanism. Finally, Section~\ref{Conclusions} provides a summary of our findings and discusses their broader implications in the context of dissipation-engineered localization.

\section{Model}\label{Model}
The dynamics of a quantum system that interacts with the environment appropriately is represented by the density matrix as the state. In this scenario, the time-evolution of the state is governed by the Lindbald master equation~\cite{breuer2002theory, lindblad1976generators}, which is described as,
\begin{equation}
    \frac{d\rho(t)}{dt} = \mathcal{L}[\rho(t)] = -i[H,\rho(t)]+\mathcal{D}[\rho(t)]
    \label{Eq:LME}
\end{equation}
where $\mathcal{L}$ is the Lindbladian superoperator, including the effects of unitary dynamics by the Hamiltonian of the quantum system and dissipation dynamics by the environmental effects. The dissipation part of the Lindbladian equation is given by,
\begin{equation}
    \mathcal{D}[\rho(t)] = \gamma\sum_{n} (S_n \rho(t) S^{\dagger}_n - \frac{1}{2} \{S^{\dagger}_n S_n, \rho(t)\} )
    \label{Eq:Dissipation_part}
\end{equation}
where $S_n$ are the jump operators, representing the dissipation channels indexed by $n$. We choose the dissipation rate $\gamma$ to be uniform across all channels. Due to the time-independent Hamiltonian ($H$) and dissipation channels ($S_n$), the Lindbladian $\mathcal{L}$ remains constant in time. Consequently, the time evolution of the density matrix can be written as,
\begin{equation}
    \rho(t) = e^{\mathcal{L}t} \rho(0)
    \label{Eq:Time_evolution_L}
\end{equation}
where the initial state is given by $\rho(0)$.
The steady-state is achieved at long-times ($t\rightarrow\infty$), satisfying,
\begin{equation}
    \rho_{ss} = \lim_{t\rightarrow \infty} \rho(t) .
    \label{Eq:Steady_state}
\end{equation}
In the Lindbladian framework, the steady-state is the eigenstate of the Lindbladian with an eigenvalue of zero: $\mathcal{L}[\rho_{ss}]=0$~\cite{alicki2007quantum}. Such steady-states will be of our key interest in this work.

We consider a single-particle tight-binding lattice system in one dimension~\cite{ashcroft1976solid}. The clean Hamiltonian of our model with dissipation is given by,
\begin{equation}
H = t \sum_{n=1}^{N-1}  (c^{\dagger}_{n} c_{n+1} + h.c.)
\label{Eq:Hamiltonian}
\end{equation}
where $c_{n}$ and $c^{\dagger}_{n}$ are the annihilation and creation operators at site $n$, respectively, and $N$ denotes the system size. The parameter $t$ represents the uniform nearest-neighbor hopping strength in the lattice. Due to the translational invariance of the Hamiltonian, the eigenstates are extended plane waves (labeled  by the wave vector $k$), and the eigenvalues are given by,
\begin{equation} 
\label{Eq:Energy_spectrum}
E_k = 2t \cos(k), \quad k = \frac{2\pi m}{N}, m= -N/2, ...,N/2
\end{equation}
under periodic boundary conditions.
Throughout this study, we use dimensionless units by setting $t=1$ and $\gamma=1$ in all numerical simulations.

Anderson localization (AL) arises due to destructive interference among multiple scattering paths in a disordered potential~\cite{lagendijk2009fifty}. This interference mechanism suppresses transport and leads to the spatial confinement of eigenstates in real space. Anderson localization relies on the presence of quantum coherence, which preserves phase information, and on destructive interference between different scattering paths that suppresses transport. However, in open quantum systems, dissipation typically breaks coherence and suppresses interference, thereby preventing localization and promoting delocalization. In this work, we propose a dissipation-driven mechanism that replicates the essential features of AL, namely, coherence and interference, without requiring any disorder in the Hamiltonian.  Evidently distinct from disorder-driven localization (AL), our approach relies on engineered dissipation to maintain coherence and enable interference, leading to steady-state confinement similar to disordered-induced localization.

With our motivation elaborated above, we adopt a phase-modulated
dissipative operator previously shown to stabilize specific eigenstates in open quantum systems~\cite{yusipov2017localization, liu2024dissipation, marcos2012photon, vershinina2017control}
\begin{equation}
S_{n} = (c^{\dagger}_{n} + e^{i\alpha_n} c^{\dagger}_{n+1} ) (c_{n} - e^{-i\alpha_n} c_{n+1} ).
\label{Eq:Dissipator}
\end{equation}
This operator acts on nearest-neighbor lattice sites with a tunable phase parameter $\alpha_n$, which controls the interference structure induced by the dissipation. Earlier work~\cite{vershinina2017control} demonstrated that, when $\alpha_n=\alpha$ is uniform across all sites, the dissipator supports a dark state at wavevector $k=\alpha$, which remains unaffected by the dissipative dynamics. In this case, the plane wave corresponding to that wavevector $k$ is annihilated by all local dissipators $S_{n}$, and thus remains unaffected by the dissipation, becoming dominant in the steady-state. If $\alpha$ does not coincide with any allowed momentum eigenvalue, then no exact dark state exists. However, the plane waves closest to $\alpha$ in momentum space experience the least dissipation and thus become dominant in the steady-state. 
% For example, When $\alpha=0$, the dissipation selectively stabilizes symmetric in-phase modes, effectively synchronizing neighboring sites by enforcing a uniform phase relationship between them~\cite{yusipov2017localization, liu2024dissipation}.

Given the ability of uniform dissipation to selectively stabilize momentum eigenstates, we now introduce a spatially varying phase profile in the dissipation operator. This allows us to explore whether dissipation can induce localization through interference alone. Specifically, as the main distinction from early work, we now demand or design the phase term in the dissipation operator as the following~\cite{sarma1988mobility, sarma1990localization},
\begin{equation}
\label{Eq:phase_site}
\alpha_n = \alpha_0 + \alpha_1 \cos(2\pi \beta n^\nu),
\end{equation}
where $\alpha_0$ is the uniform initial phase across all dissipation channels, and $\alpha_1$ controls the modulation strength.  The exponent $\nu$ governs the rate of spatial variation, ensuring a gradual modulation across the lattice when $0<\nu<1$. This condition introduces aperiodicity in the dissipation mechanism, as the phase modulation follows a slowly varying function of $n^{\nu}$. This slow variation is essential as it introduces long-range phase correlations across the lattice, allowing coherence to build up and enabling interference even in the absence of disorder.
The parameter $\beta$ determines whether this aperiodic modulation is commensurate or incommensurate, depending on whether it takes a rational or irrational value~\cite{sarma1988mobility}. We refer to cases with rational $\beta$ as "aperiodic commensurate" and those with irrational $\beta$ as "aperiodic incommensurate". Despite this distinction, the modulation remains aperiodic in both cases due to the nonlinear scaling of $n^{\nu}$.  However, as we will see in our results, commensurate and incommensurate $\beta$ values do not lead to identical localization behavior, revealing remarkable differences in their steady-state properties.

To characterize the steady-state, we employ three key observables: the relative entropy of coherence {$C_{\rm re}$~\cite{baumgratz2014quantifying}}, the Purity~\cite{nielsen2010quantum}, and the participation ratio (PR)~\cite{longhi2024dephasing}. These measures provide complementary insights into the coherence, mixedness, and localization properties of the steady-state.

The relative entropy of coherence, $C_{\rm{re}}$, quantifies the quantum coherence of the steady-state by measuring how much the density matrix $\rho_{ss}$ deviates from a fully incoherent state~\cite{baumgratz2014quantifying} {in a chosen representation (the diagonal basis)}. It is defined as,
\begin{equation} 
C_{\rm{re}} = S(\rho_{ss,diag}) - S(\rho_{ss}), 
\label{Eq:C_re}
\end{equation}
where $S(\rho) = -\rm{Tr}(\rho ~{\rm{log}}(\rho))$ is the von Neumann entropy, and $\rho_{ss,diag}$ is the diagonal part of $\rho_{ss}$ in a chosen basis. A larger $C_{\rm{re}}$ indicates greater coherence, while $C_{\rm{re}}=0$ corresponds to a fully incoherent state. This measure helps assess how dissipation affects the coherence properties of the steady-state. 

The purity of the steady-state density matrix quantifies how mixed or pure the system is~\cite{nielsen2010quantum}. It is defined as,
\begin{equation} 
\text{\rm Purity} = \text{Tr}(\rho_{\text{ss}}^2).
\label{Eq:purity}
\end{equation}
A pure state corresponds to 
$\rm{Purity}=1$ while a maximally mixed state has $\rm{Purity}=1/N$. Purity serves as a measure of the impact of dissipation, with lower values indicating an increased decoherence and a more mixed steady-state.

The participation ratio (PR) quantifies the number of lattice sites that significantly contribute to the steady-state density matrix in the site basis~\cite{longhi2024dephasing}. It is defined as,
\begin{equation} \text{PR} = \frac{1}{\sum_n [\rho_{ss}]_{nn}^2}. 
\label{Eq:PR}
\end{equation}
A smaller PR indicates that the steady-state population is concentrated in a fewer sites, signifying localization, while a larger PR suggests a more delocalized steady-state nature. Thus, PR serves as a key measure for assessing whether dissipation leads to localization or delocalization.

Having established the system Hamiltonian (eq.~\ref{Eq:Hamiltonian}) and the phase-modulated dissipation operator (eq.~\ref{Eq:Dissipator}), we now examine their impact on the steady-state properties. Specifically, we investigate whether engineered aperiodically modulated dissipation can induce localization even when the Hamiltonian eigenstates remain extended. The following section presents our detailed results and key observations, demonstrating the emergence of dissipation-induced localization in an otherwise clean and extended system.

%%%%%%%%%%%%%%%%%%%%%%%%%%%%%%%%%%%%%%%%%%%%%%%%%%%%%%%%%%%%%%
\section{Results:}\label{Results}
\begin{figure}[t!]

    \includegraphics[width=1\columnwidth]{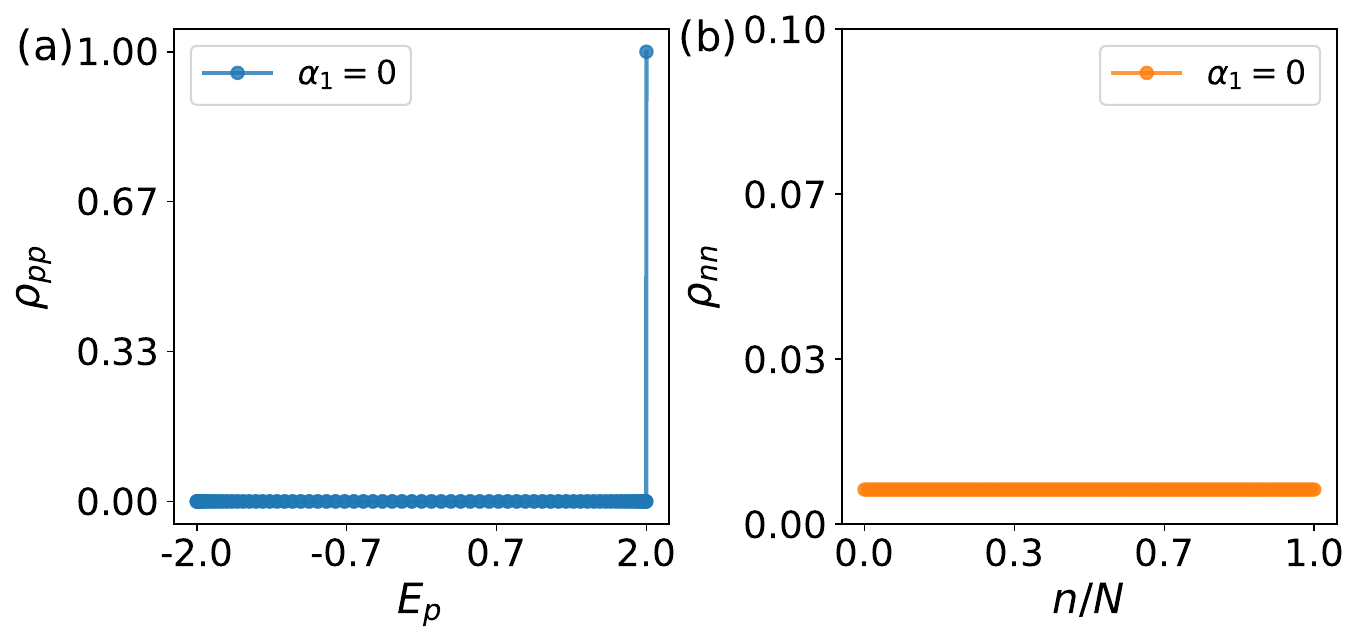}
    \caption{(a) Steady-state population in the eigenbasis ($\rho_{pp}$) as a function of the eigenenergies ($E_p$) of the Hamiltonian~[eq.(\ref{Eq:Hamiltonian})]. (b)   Steady-state population in the site basis ($\rho_{nn}$) as a function of the normalized site index ($n/N$). Here the results are shown for a system size of $N=144$ with $\alpha_0 = 0$. 
    \label{Fig:Fig1_QP}}
\end{figure}

%%%%%%%%%%%%%%%%%%%%%%%%%%%%%%%%%%%%%%%%%%%%%%%%%%%%%%%%%%%%%%
To begin with, we set $\alpha_0 = 0$, ensuring that dissipation operator acts uniformly across all lattice sites. As a validation of the known results~\cite{vershinina2017control} of the model, we confirm that the steady-state is a pure state; as shown in Fig.~\ref{Fig:Fig1_QP}(a), where we plot the steady-state population in the eigenbasis, defined as $\rho_{pp}=\langle p|\rho_{ss}|p\rangle$. Similarly, Fig.~\ref{Fig:Fig1_QP}(b) presents the steady-state population in the site basis $\rho_{nn}=\langle n|\rho_{ss}|n\rangle$, which provides information on the spatial distribution of the steady-state. 
This behavior can be understood by analyzing the energy spectrum of the Hamiltonian (eq.~\ref{Eq:Energy_spectrum}). A single eigenstate remains protected from dissipation if it satisfies the dark state condition $S_{n}|p\rangle=0$ for all dissipation channels (eq.~\ref{Eq:Dissipator}). This occures when the dissipation phase parameter $\alpha_0$ coincides with the wavevector $k$ of the eigenstate in the momentum basis~\cite{vershinina2017control}. Since the highest-energy eigenstate corresponds to $k=0$, it naturally survives as the steady-state. Due to its uniform plane wave structure, this eigenstate is extended across the entire lattice. As a result, the steady-state population in the site basis $\rho_{nn}$ remains delocalized, as shown in ~Fig.~\ref{Fig:Fig1_QP}(b).

Using the previously discussed set up as a reference, we now introduce site-dependent dissipation by modulating the phase parameter $\alpha_n$ in eq.~\ref{Eq:phase_site}. This modulation introduces three key parameters: the modulation strength $\alpha_1$, the modulation periodicity $\beta$, and the spatial variation rate $\nu$. To systematically understand their impact, we first analyze the effects of $\beta$ that is, incommensurate~[\ref{Incommensurate_case}] and commensurate~[\ref{Commensurate_case}] separately. For each case, we explore how $\alpha_1$ and $\nu$ influence the steady-state properties and compare their effects across incommensurate and commensurate cases.

\subsection{Aperiodic Incommensurate Case}{\label{Incommensurate_case}}
We now analyze the impact of site-dependent dissipation, introduced via an aperiodic incommensurate phase modulation in $\alpha_n$ (eq.~\ref{Eq:phase_site}). Unlike the uniform dissipation case ($\alpha_0 = 0$, Fig.~\ref{Fig:Fig1_QP}) where the steady-state remains delocalized, an incommensurate $\alpha_n$ introduces a nontrivial interplay between dissipation and Hamiltonian eigenstates. To systematically characterize this effect, we examine the steady-state behavior in both the site basis ($\rho_{nm}$) and in the eigenbasis ($\rho_{pq}$). Throughout this section, we set $\beta$ to the inverse golden ratio, that is, $\beta = \frac{\sqrt{5}-1}{2}$, ensuring a strong aperiodic incommensurate modulation with long-range correlations~\cite{sarma1988mobility}.

\begin{figure}[t!]

    \includegraphics[width=1\columnwidth]{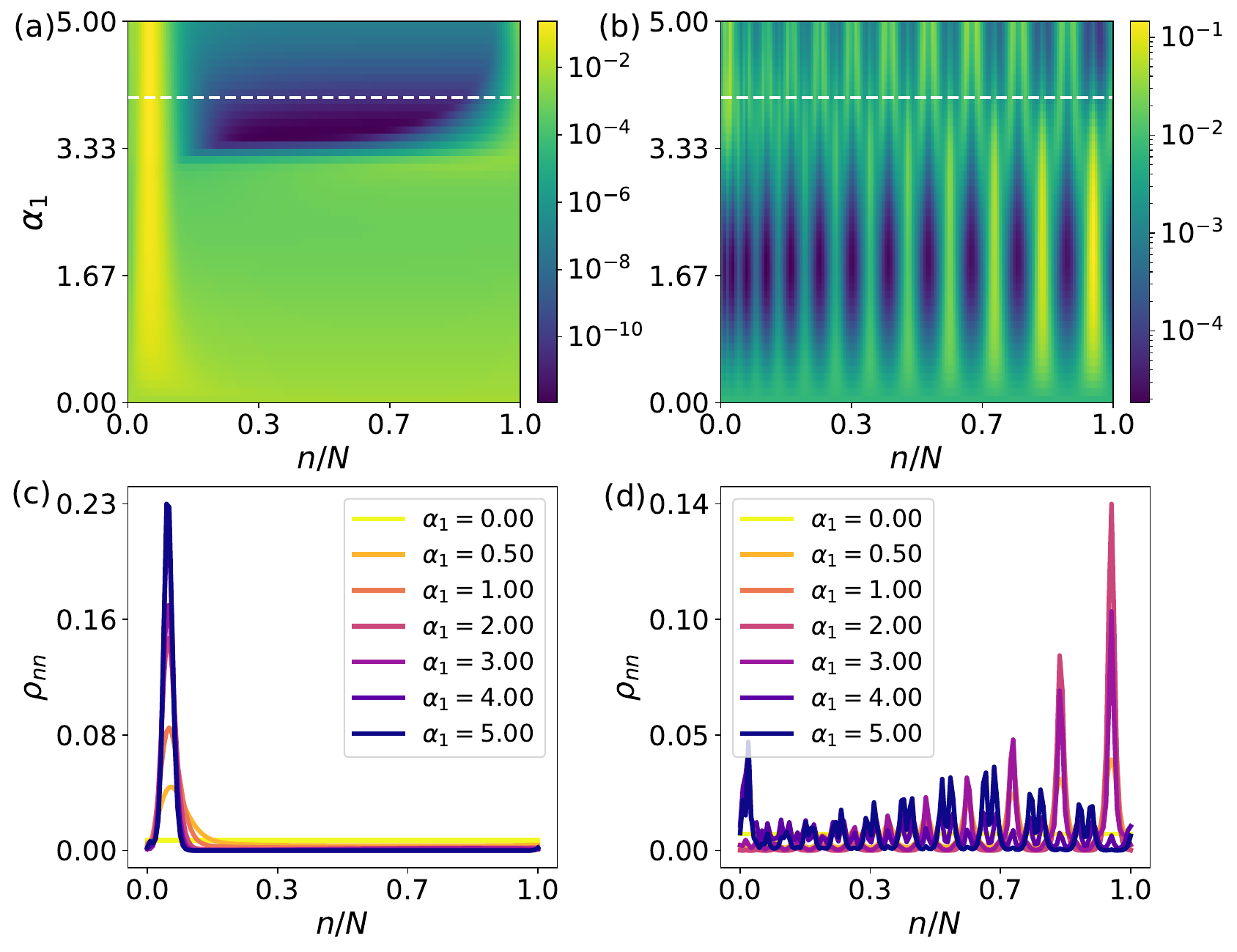}
    \caption{Incommensurate case: Steady-state population in the site basis ($\rho_{nn}$ ) plotted in the $\alpha_1$-$n/N$ plane for (a) $\nu = 0.1$ and (b) $\nu=0.6$, where the color scale represents log$\rho_{nn}$. The white line at $\alpha_1=4$ indicates the parameter choice for further analysis. Panels (c) and (d) show $\rho_{nn}$ as a function of $n/N$ for selected values of $\alpha_1$ corresponding to (a) and (b), respectively. All calculations are performed for a system size of $N=144$.  
    \label{Fig:Fig2_QP}}
\end{figure}

To understand how the modulated dissipation affects the real-space distribution of the steady-state, we analyze the population in the site basis ($\rho_{nn}$) as a function of the modulation strength $\alpha_1$ and the normalized site index $n/N$ as shown in Fig.~\ref{Fig:Fig2_QP}. Panels (a) and (b) compare two representative values of the spatial variation rate $\nu$, which controls how rapidly the dissipative phase $\alpha_n$ changes across the lattice. For $\nu=0.1$ (slow modulation), the dissipation phase varies gradually along the lattice inducing long-range phase correlations. In contrast, for $\nu=0.6$ (fast modulation), it changes more rapidly between sites, resembling random dephasing. The color scale represents log$\rho_{nn}$, illustrating how the dissipation-induced population distribution evolves along the lattice sites.
In the slow modulation regime ($\nu=0.1$, Fig.~\ref{Fig:Fig2_QP}(a))
, the steady-state population remains extended across a significant portion of the lattice for small $\alpha_1$, consistent with delocalized behaviour. However, as $\alpha_1$ approaches $\pi$, the population becomes concentrated in a small number of sites; we will refer to this as localization, hence signaling the emergence of dissipation-induced localization. To further illustrate this behavior, Fig.~\ref{Fig:Fig2_QP}(c) presents steady-state population distributions for representative values of $\alpha_1$. At low $\alpha_1$, the population is spread relatively evenly across the system, whereas for larger $\alpha_1$, the population is increasingly concentrated at specific sites. This observation demonstrates how the population gradually shifts from a delocalized to a more localized configuration.
In contrast, for the fast modulation regime ($\nu=0.6$, Fig.~\ref{Fig:Fig2_QP}(b)), the steady-state population remains extended across the lattice, exhibiting no significant localization across the full range of $\alpha_1$. This suggests that a higher spatial variation rate prevents steady-state localization. A more detailed view of the population distributions for various $\alpha_1$ values provided in Fig.~\ref{Fig:Fig2_QP}(d), further confirming that faster phase variations suppress localization effects entirely.

To facilitate further analysis, we highlight a "white line" at $\alpha_1 = 4$ in Fig.~\ref{Fig:Fig2_QP}(a, b), marking the modulation strength chosen for further analysis. This particular value is selected because it represents a regime where localization effects become pronounced, allowing for a detailed study of steady-state localization properties.

\begin{figure}[t!]

    \includegraphics[width=1\columnwidth]{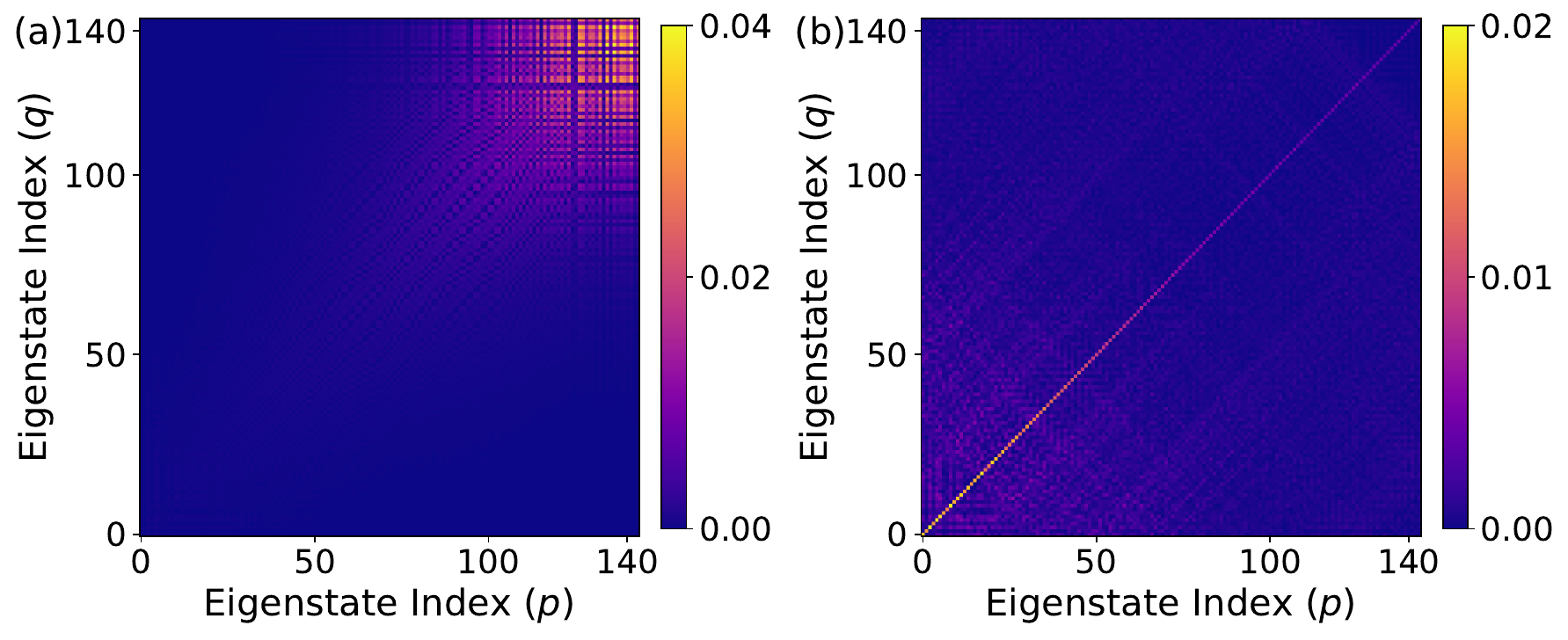}
    \caption{Incommensurate case: Heatmaps of the steady-state density matrix in the eigenbasis ($\rho_{pq}$) for (a) $\nu = 0.1$ and (b) $\nu=0.6$, illustrating the emergence and absence of steady-state coherence, respectively. This analysis is performed at $\alpha_1 = 4$, corresponding to the white line in Fig~.\ref{Fig:Fig2_QP}(a) and (b).  The system size for these calculations is $N= 144$. 
    \label{Fig:Fig3_QP}}
\end{figure}

To gain qualitative insight into the mechanism underlying this dissipation-induced localization, we analyze the steady-state density matrix in the eigenbasis. Fig.~\ref{Fig:Fig3_QP}(a, b) demonstrate $\rho_{pq}$ for (a) $\nu=0.1$ and (b) $\nu=0.6$, both at a modulation strength of $\alpha_1 = 4$.
For the slow modulation case ($\nu=0.1$), Fig.~\ref{Fig:Fig3_QP}(a) shows that the population is predominantly concentrated in specifically favored eigenstates, with noticeable off-diagonal elements in $\rho_{pq}$. These off-diagonal contributions indicate that dissipation does not simply generate an incoherent statistical mixture of eigenstates. Instead, it induces a coherent superposition of eigenstates, leading to persistent interference effects in real-space at $\alpha_1=4$. This non-trivial interplay between dissipation and eigenstates nature effectively constrains the steady-state population to spatial regions, resulting in real-space localization, as observed in Fig.~\ref{Fig:Fig2_QP}(a, c).
In contrast, for the fast modulation case ($\nu=0.6$), Fig.~\ref{Fig:Fig3_QP}(b) reveals predominantly diagonal steady-state density matrix, indicating that dissipation acts as a decoherence mechanism, reducing coherence between different energy eigenstates, analogous to the effect of strong dephasing. As a result, the steady-state population remains extended in real space, consistent with the delocalized behaviour observed in Fig.~\ref{Fig:Fig2_QP}(b, d). This confirms that faster phase variations suppress localization effects by inhibiting dissipation-induced eigenstates coupling and promoting classical statistical mixture.

These results indicate that at $\alpha_1=4$, dissipation induces a non-trivial coherent superposition of momentum states in the slow modulation regime ($\nu=0.1$), leading to real space localization. Conversely, in the fast modulation regime ($\nu=0.6$), the eigenstate population remains broadly distributed, forming an incoherent statistical mixture that prevents localization, resulting in delocalization in real space.

\begin{figure}[t!]

    \includegraphics[width=1\columnwidth]{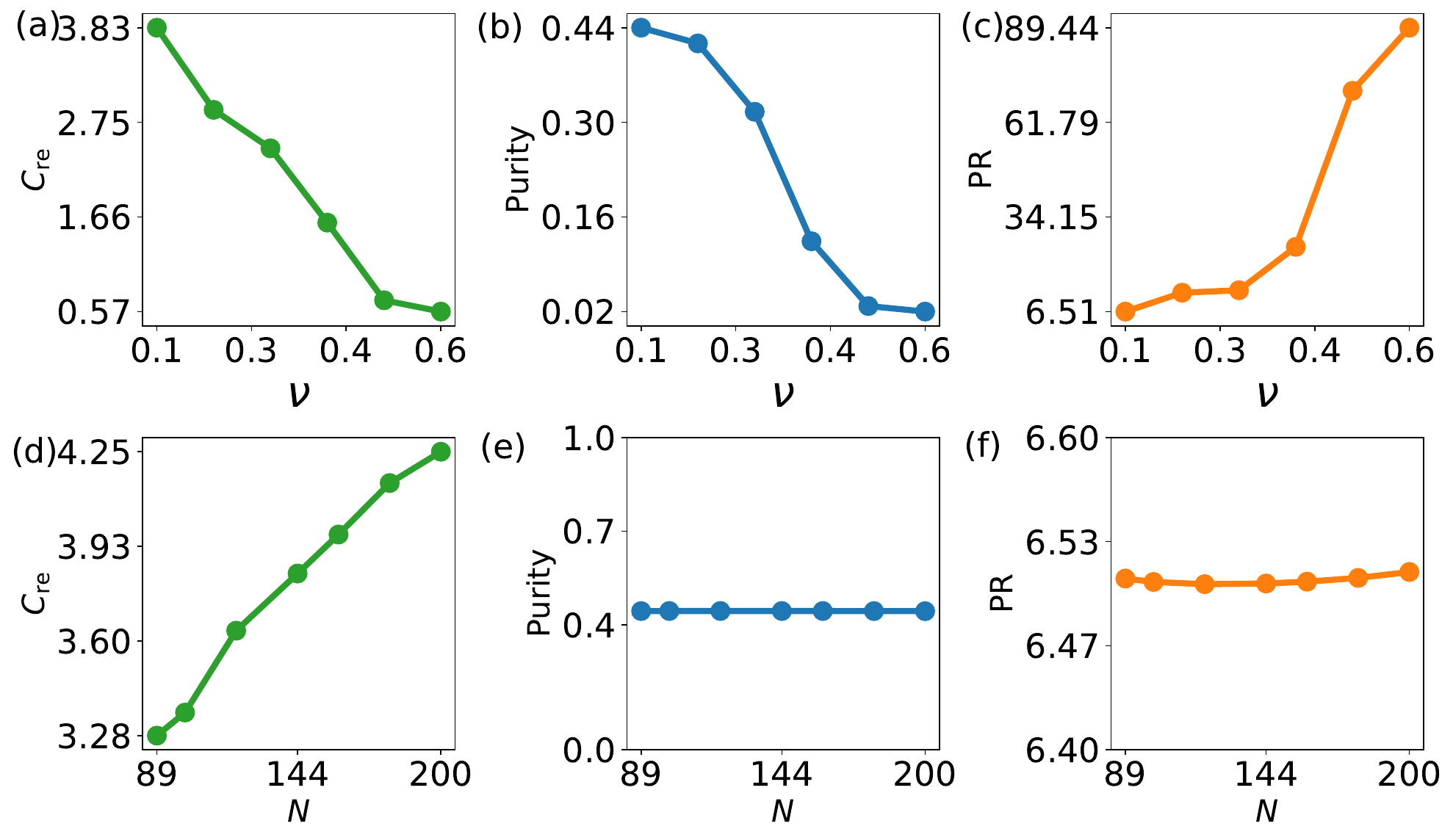}
    \caption{Incommensurate case: (a) Relative entropy of coherence ($C_{\rm{re}}$) (b) Purity (c) Participation ratio (PR) as functions of $\nu$ for $\alpha_1 = 4$ and $N=144$. Panels (d), (e), and (f) show the system-size dependence of these quantities for $\nu=0.1$. 
    \label{Fig:Fig4_QP}}
\end{figure}

To investigate the interplay between coherence in the eigenbasis and localization in the site basis, we analyze three key observables: relative entropy of coherence $C_{\rm{re}}$ (eq.~\ref{Eq:C_re}), Purity (eq.~\ref{Eq:purity}), and participation ratio (PR) (eq.~\ref{Eq:PR}) as functions of the spatial variation rate $\nu$, shown in Fig.~\ref{Fig:Fig4_QP}(a-c). Since $C_{\rm{re}}$ and Purity quantify coherence in the eigenbasis, while PR measures localization in the site basis, their combined analysis allows us to determine whether steady-state localization is directly correlated with the degree of coherence maintained in the eigenbasis.
Fig.~\ref{Fig:Fig4_QP}(a) shows that $C_{\rm{re}}$ decreases significantly with increasing $\nu$, indicating that faster phase modulation suppresses coherence in the steady-state. Specifically, $C_{\rm{re}}$ drops from $3.83$ at $\nu=0.1$ to $0.57$ at $\nu =0.6$, highlighting a significant trend of reduction in coherence as the modulation rate increases. 
Similarly, Fig.~\ref{Fig:Fig4_QP}(b) reveals a decline in Purity, confirming that the steady-state becomes increasingly mixed as $\nu$ increases. Purity decreases from $0.44$ at $\nu=0.1$ to $0.02$ at $\nu=0.6$, indicating that the steady-state approaches as a nearly fully mixed state at high modulation rates. Notably, both these quantities, $C_{\rm{re}}$ and Purity exhibits saturation behavior for $\nu\geq 0.5$, suggesting that beyond this threshold, further increases in the modulation rate do not significantly alter the coherence properties of the steady-state.
Fig.~\ref{Fig:Fig4_QP}(c) presents the participation ratio (PR), which quantifies the extent of localization in the site basis. At $\nu=0.1$, PR is $6.51$, indicating that the steady-state population is concentrated in a few sites, signaling a clear signature of localization. As $\nu$ increases, PR rises gradually until $\nu=0.4$, after which it exhibits a sharp increase, reaching at $71$ at $\nu=0.5$ and $89.44$ at $\nu=0.6$. This trend confirms that faster modulation enhances delocalization, with steady-state population significantly distributed across many sites beyond $\nu=0.4$.
Collectively, these results demonstrate that increasing the spatial variation rate $\nu$ enhances decoherence and mixedness in the eigenbasis, while simultaneously promoting delocalization in the real space. This support the conclusion that dissipation-induced localization critically depends on the preservation of the long-range phase correlations, which is disrupted at fast modulation rates.

The system size dependence results in Fig.~\ref{Fig:Fig4_QP}(d–f) confirm the robustness of the observed trends. Here, we focus on the case $\nu=0.1$ to validate the persistence of dissipation-induced effects across different system sizes. For this analysis, we consider $N= 89, 100, 120, 144, 160, 180$, and $200$.
Fig.~\ref{Fig:Fig4_QP}(d) shows that $C_{\rm{re}}$ remains finite as the system size increases, indicating the steady-state retains coherence across a growing number of eigenstates. This suggests that dissipation does not fully suppress eigenstate coherence but instead allows for coherent superposition of energy eigenstates even in larger systems.
Similarly, Fig.~\ref{Fig:Fig4_QP}(e) shows that Purity remains nearly unchanged across all system sizes, indicating that the degree of mixedness in the steady-state is not a system-size effect but rather an intrinsic property of the dissipation dynamics.
Finally, Fig.~\ref{Fig:Fig4_QP}(f) confirms that PR remains stable with increasing $N$, demonstrating that the steady-state population remains localized within a finite number of sites, independent of system size. 
Together, these results demonstrate that the localization is robust and persists in the thermodynamic limit. 

These findings establish a clear and physically intuitive connection between coherence in the eigenbasis and localization in the real space: greater coherence corresponds to stronger localization (low PR), while weaker coherence leads to delocalization (high PR). Notably, this dissipation-induced localization emerges in the slow modulation regime, where long-range phase correlations enable the formation of non-trivial coherent superpositions of momentum eigenstates. These superpositions interfere constructively in real space, producing localized steady-state population distribution. In contrast, in the faster modulation regime, $\alpha_n$ fluctuates rapidly across neighboring sites, introducing random-like, short-range phase variation. This rapid variation disrupts phase coherence and mimics a local dephasing mechanism that suppresses interference between eigenstates. As a result, the steady state evolves into an incoherent statistical mixture of delocalized momentum eigenstates, thereby destroying localization. Thus, faster modulation inhibits the interference effects required to sustain dissipation-induced localization.

Having demonstrated that aperiodic incommensurate modulation ($\beta = \frac{\sqrt{5}-1}{2}$) leads to localization for slow modulation ($\nu=0.1$) and delocalization for fast modulation ($\nu=0.6$), a natural question arises:
Does this behavior persist if the phase modulation follows an aperiodic commensurate pattern?

\begin{figure}[t!]

    \includegraphics[width=1\columnwidth]{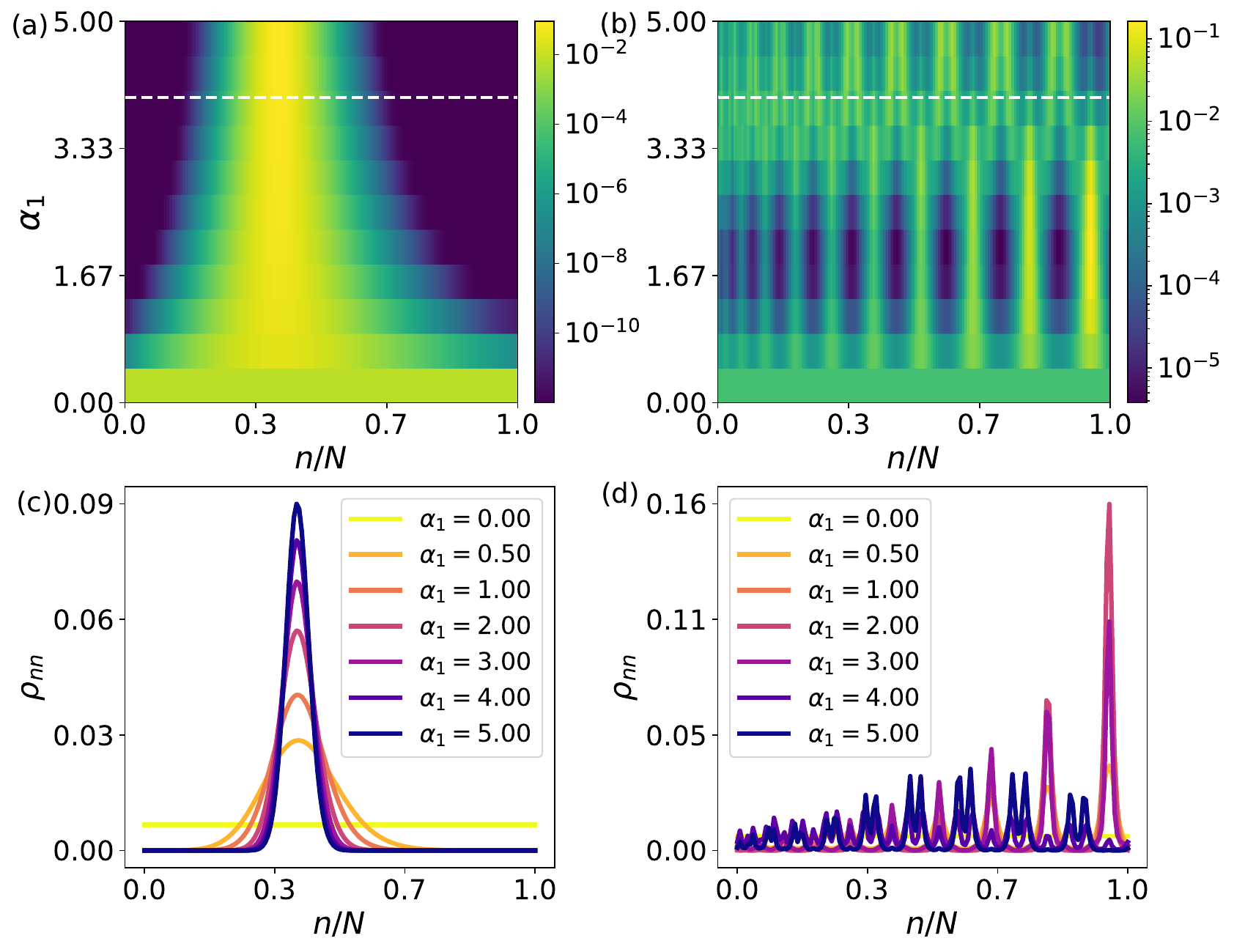}
    \caption{Commensurate case: Steady-state population in the site basis ($\rho_{nn}$ ) plotted in the $\alpha_1$-$n/N$ plane for (a) $\nu = 0.1$ and (b) $\nu=0.6$, where the color scale represents log$\rho_{nn}$. The white line at $\alpha_1=4$ indicates the parameter choice for further analysis. Panels (c) and (d) show $\rho_{nn}$ as a function of $n/N$ for selected values of $\alpha_1$ corresponding to (a) and (b), respectively. All calculations are performed for a system size of $N=150$ and $\beta = 0.5$. 
    \label{Fig:Fig1_periodic}}
\end{figure}

\subsection{Aperiodic Commensurate Case}{\label{Commensurate_case}}
While the aperiodic incommensurate case, $\beta = \frac{\sqrt{5}-1}{2}$ led to strong dissipation-induced localization for slow modulation and delocalization for fast modulation, it remains unclear whether this behavior is a generic consequence of aperiodic phase modulation introduced by $n^{\nu}$ or if it fundamentally depends on whether $\beta$ is rational or irrational. To address this question, we now consider an aperiodic commensurate phase by setting $\beta=0.5$, allowing us to assess how the nature of $\beta$ influences the steady-state properties. By systematically analyzing the steady-state properties under this  modulation, we investigate whether similar localization and coherence effects emerge, or if the distinction between rational and irrational $\beta$ plays an essential role in inducing localization.

Now, we examine the same key observables as in the incommensurate case, but focus on their behavior under aperiodic commensurate modulation. Fig.~\ref{Fig:Fig1_periodic} presents the steady-state population ($\rho_{nn}$) distribution in the site basis as a function of the modulation strength $\alpha_1$ and normalized site index $n/N$ allowing for a direct comparison with Fig.~\ref{Fig:Fig2_QP}.
Unlike the incommensurate case, where slow modulation ($\nu=0.1$) led to strong localization, commensurate modulation does not exhibit clear localization effects even at high modulation strength $\alpha_1$. Fig.~\ref{Fig:Fig1_periodic}(a) and Fig.~\ref{Fig:Fig1_periodic}(b) compare the steady-state population distribution for slow ($\nu=0.1$) and fast ($\nu=0.6$) modulation, respectively.
In the case of slow modulation [Fig.~\ref{Fig:Fig1_periodic}(a)], at large $\alpha_1$ the population remains more broadly distributed across the system as compared to the incommensurate case (Fig.~\ref{Fig:Fig2_QP}(a, c)).
On the other hand, for the fast modulation [Fig.~\ref{Fig:Fig1_periodic}(b)], the steady-state remains delocalized, similar to the behavior observed in the incommensurate case at $\nu=0.6$ (Fig.~\ref{Fig:Fig2_QP}(b, d)). Fig.~\ref{Fig:Fig1_periodic}(c, d) show line cuts of the population distribution for selected values of $\alpha_1$ . Unlike in the incommensurate case, where increasing $\alpha_1$ led to sharply localized peak, here the population remains relatively smooth and broadly distributed across sites. 
These results suggests that while both cases break translational symmetry, the incommensurate case leads to strong localization, whereas the commensurate case results in a more extended but still weakly confined steady-state.

 \begin{figure}[t!]

    \includegraphics[width=1\columnwidth]{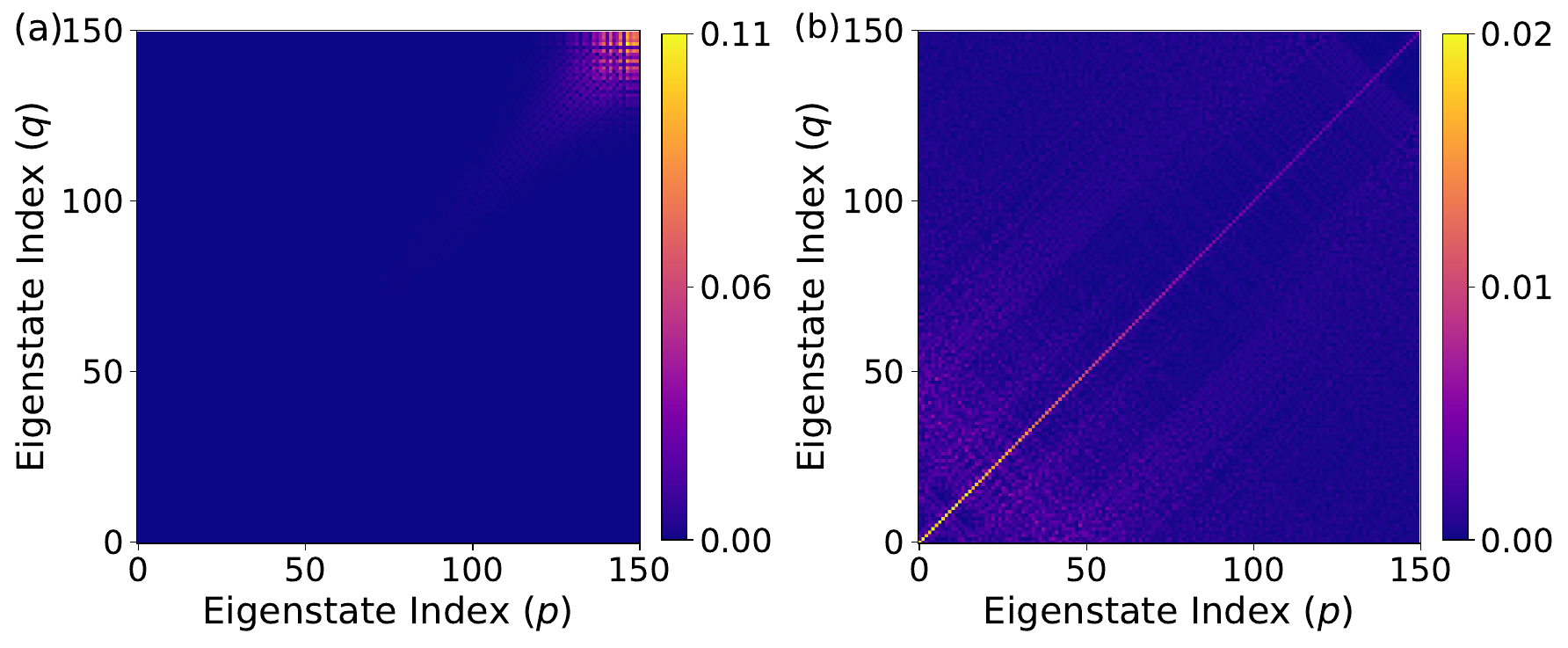}
    \caption{Commensurate case: Heatmaps of the steady-state density matrix in the eigenbasis ($\rho_{pq}$) for (a) $\nu = 0.1$ and (b) $\nu=0.6$, illustrating the emergence and absence of steady-state coherence, respectively. This analysis is performed at $\alpha_1 = 4$, corresponding to the white line in Fig.~\ref{Fig:Fig1_periodic}(a) and (b).  The system size for these calculations is $N=150$ and $\beta = 0.5$.  
    \label{Fig:Fig2_periodic}}
\end{figure}

To understand why aperiodic commensurate modulation does not induce strong localization, we analyze the steady-state density matrix in the eigenbasis ($\rho_{pq}$), as shown in Fig.~\ref{Fig:Fig2_periodic}. Fig.~\ref{Fig:Fig2_periodic}(a, b) present $\rho_{pq}$ for slow ($\nu=0.1$) and fast $\nu=0.6$ modulation, respectively.
In the case of slow modulation [Fig.~\ref{Fig:Fig2_periodic}(a)], the population is distributed across selective eigenstates, with only weak off-diagonal coherence. This sharply contrasts with the incommensurate case, where slow modulation preferentially populated significant number of eigenstates and preserved coherence, ultimately leading to localization. The absence of strong coherence signatures in the commensurate case suggests that choosing commensurate $\beta$ reduces the effects of long-range phase correlations, weakening the conditions necessary for localization. 
For the fast modulation case [Fig.~\ref{Fig:Fig2_periodic}(b)], $\rho_{pq}$ becomes nearly diagonal, indicating that the system reaches a fully incoherent mixture of extended eigenstates. This behavior is similar to the incommensurate case at high $\nu$, confirming the dephasing mechanism.

\begin{figure}[t!]

    \includegraphics[width=1\columnwidth]{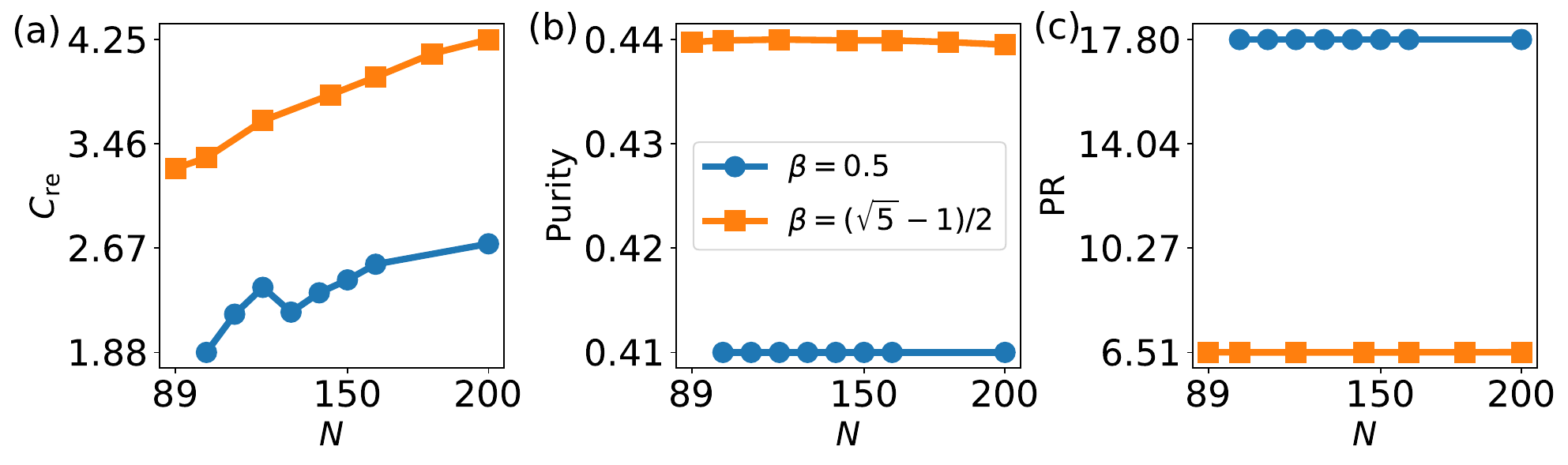}
    \caption{Incommensurate and Commensurate comparison: System-size dependence of steady-state properties for commensurate ($\beta=0.5$, square symbols) and incommensurate ($(\sqrt 5-1)/2$, circle symbols) modulations. (a) Relative entropy of coherence $C_{\rm{re}}$, (b) Purity, and (c) participation ratio (PR) are shown as functions of system size $N$. All calculations are performed at $\alpha_1 = 4$ and $\nu=0.1$. The comparison reveals distinct differences in coherence scaling and localization behavior between the two modulation types.
    \label{Fig:Fig3_periodic}}
\end{figure}

Fig.~\ref{Fig:Fig3_periodic} presents a system size dependence of coherence $C_{\rm{re}}$, Purity, and participation ratio (PR) for both incommensurate and commensurate modulation, allowing for a direct comparison of their effects on dissipation-induced localization. All calculations are performed for $\nu=0.1$ (slow modulation) and $\alpha_1 = 4$.
Fig.~\ref{Fig:Fig3_periodic}(a) shows the evolution of relative entropy of coherence $C_{\rm{re}}$ as a function of $N$. In the commensurate case, $C_{\rm{re}}$ increases from $1.88$ to $2.7$, whereas in the incommensurate case, it increases from $3.35$ to $4.25$. This difference highlights that although coherence grows with system size in both cases, incommensurate modulation sustains significantly higher coherence levels.
Fig.~\ref{Fig:Fig3_periodic}(b), examines Purity, which remains nearly unchanged across system sizes in both cases. This confirms that the degree of mixedness in the steady-state is not strongly affected by the system sizes.
Finally, Fig.~\ref{Fig:Fig3_periodic}(c), displays the participation Ratio (PR) as a function of $N$. In the commensurate case, PR remains constant at $17.8$, whereas in the incommensurate case, PR remains fixed at a much lower value of $6.51$. This result directly reflects the stronger localization effect in the incommensurate case, where the steady-state population is confined to fewer sites. In contrast, commensurate modulation leads to a weakly localized steady-state. This result concludes that rational values of $\beta$ inhibit the buildup of long-range phase correlations, which are essential for supporting coherent interference and achieving strong dissipation-induced localization.

\section{Experimental proposal:}\label{Experimental_Proposal}
To explore the experimental feasibility of our proposed dissipation mechanism (eq.~\ref{Eq:Dissipator}), we outline a possible implementation strategy inspired by Ref.~\cite{marcos2012photon}. The setup involves a circuit quantum electrodynamics (QED) architecture consisting of two superconducting microwave cavities, each coupled to a superconducting qubit. These cavities support symmetric and antisymmetric modes defined as  $ c_s = (c_1 + c_2)/\sqrt{2} $  and $c_a = (c_1 - c_2)/\sqrt{2}$ , respectively. In the original scheme, each qubit is driven by a classical microwave field. The drive amplitudes are equal in magnitude but opposite in phase, given by $\Omega$  and $-\Omega$. This configuration selectively enables transitions that couple to the antisymmetric mode $c_a$. After these transitions, the qubits emit photons spontaneously, which causes excitations to move irreversibly from the antisymmetric mode ($c_a$) to the symmetric mode ($c_s$). By adiabatically eliminating the qubits, this irreversible process can be described by an effective dissipator of Lindblad form $\mathcal{D}[c_s^\dagger c_a]$ . Specifically, the resulting jump operator ($S = c_s^\dagger c_a$) takes the structure of a product between creation and annihilation operators, where the antisymmetric mode is coherently absorbed and the symmetric mode is re-emitted. This matches the form of our phase-modulated dissipation operator in eq.~\ref{Eq:Dissipator}. While the original scheme focuses on a single block with two cavities, Ref.~\cite{marcos2012photon} notes that the generalization to an extended array is straightforward. We adopt this perspective and implement a phase profile $\alpha_n$ across each bond $(n, n+1)$ in the lattice. This modification allows us to selectively couple to a generalized antisymmetric mode of the form  $( c_n - e^{-i\alpha_n} c_{n+1} )$. This can be realized by driving the two corresponding qubits with amplitudes $\Omega$  and $-\Omega e^{-i\alpha_n}$. The required spatial profile  $\alpha_n$ can be engineered using arbitrary waveform generators (AWGs)~\cite{lin2019scalable}. These devices allow programmable modulation of the microwave drives applied to the qubits. By programming the AWGs to implement modulations such as  $ \alpha_n = \alpha_0 + \alpha_1 \cos(2\pi \beta n^\nu) $ as defined in eq.~\ref{Eq:phase_site}, we construct a tunable and spatially dependent dissipation channel. This channel irreversibly transfers excitations from generalized antisymmetric modes into symmetric ones.

%%%%%%%%%%%%%%%%%%%%%%%%%%%%%%%%%%%%%%%%%%%%%%%%%%%%%%%%%%%%%%
\section{Conclusions:}\label{Conclusions}
%%%%%%%%%%%%%%%%%%%%%%%%%%%%%%%%%%%%%%%%%%%%%%%%%%%%%%%%%%
In this work, we explored how site-dependent, phase-modulated dissipation can facilitate steady-state localization in a one-dimensional tight-binding lattice system.
In particular, we considered aperiodic dissipation, which inherently induces long-range phase correlations. Our findings demonstrate that these correlations play a crucial role in determining dissipation-induced localization. We systematically compared two classes of this aperiodic dissipation (called commensurate and incommensurate cases).
Our results show that incommensurate dissipation induces strong localization, confining the steady-state population to a small number of lattice sites while preserving coherence in the eigenbasis. In contrast, commensurate dissipation fails to induce strong localization, leading to a weakly localized steady-state. The interplay between coherence, purity, and participation ratio reveals that steady-state localization in the site-basis is directly linked to coherence retention in the eigenbasis.
The system-size dependence analysis further confirms that coherence and localization persist as the system size increases, highlighting the robustness of this mechanism. 

These findings establish aperiodic dissipation as a mechanism for engineered localization, demonstrating that dissipation can be a control tool rather than just a source of decoherence. While distinct from disorder-driven AL, this mechanism produces steady-states with similar spatial confinement and coherence properties, broadening the scope of localization phenomena. By advancing our understanding of dissipation-engineered localization, these results may provide valuable insights for future applications in quantum control, decoherence suppression, and engineered quantum phases. Future studies may extend this framework to higher dimensions, interacting systems, or Floquet-driven setups, to further explore the role of dissipation in shaping quantum phases and dynamics.

%%%%%%%%%%%%%%%%%%%%%%%%%%%%%%%%%%%%%%%%%%%%%%%%%%%%%%%%%%%%%%
\begin{acknowledgements}

J.G. acknowledges support by the National Research Foundation, Singapore through the National Quantum Office, hosted in A*STAR, under its Centre for Quantum Technologies Funding Initiative (S24Q2d0009).  Authors also thank Weitao Chen and Zhixing Zou for fruitful discussions. 
\end{acknowledgements}
%%%%%%%%%%%%%%%%%%%%%%%%%%%%%%%%%%%%%%%%%%%%%%%%%%%%%%%%%%%%%%

%\section{Appendix:}

\bibliography{Ref.bib}

\end{document}